# Measurements and *ab initio* Molecular Dynamics Simulations of the High Temperature Ferroelectric Transition in Hexagonal $RMnO_3$

T. A. Tyson[1,3], T. Wu[1], H. Y. Chen[1], J. Bai[2], K. H. Ahn[1,3]
K. I. Pandya[4], S. B. Kim[3] and S.-W. Cheong[3]
[1]Department of Physics, New Jersey Institute of Technology, Newark, NJ 07102
[2]Oak Ridge National Laboratory, Oak Ridge, TN 37831
[3]Rutgers Center for Emergent Materials and Department of Physics and Astronomy, Rutgers University, Piscataway, NJ 08854
[4]Brookhaven National Laboratory, Upton New York, 11973

## Abstract

Measurements of the structure of hexagonal $RMnO_3$ (R=rare earths (Ho) and Y) for temperatures significantly above the ferroelectric transition temperature ($T_{FE}$) were conducted to determine the nature of the transition. The local and long range structural measurements were complemented by *ab initio* molecular dynamics simulations. With respect to the Mn sites in $YMnO_3$ and $HoMnO_3$, we find no large atomic (bond distances or thermal factors), electronic structure changes or rehybridization on crossing $T_{FE}$ from local structural methods. The local symmetry about the Mn sites is preserved. With respect to the local structure about the Ho sites, a reduction of the average Ho-O bond with increased temperature is found. Ab initio molecular dynamics calculations on $HoMnO_3$ reveal the detailed motions of all ions. Above ~900 K there are large displacements of the Ho, O3 and O4 ions along the z-axis which reduce the buckling of the MnO3/O4 planes. The changes result in O3/O4 ions moving to towards central points between pairs of Ho ions on the z-axis. These structural changes make the coordination of Ho sites more symmetric thus extinguishing the electric polarization. At significantly higher temperatures, rotation of the $MnO_5$ polyhedra occurs without a significant change in electric polarization. The born effective charge tensor is found to be highly anisotropic at the O sites but does not change appreciably at high temperatures.

PACS: 77.80.-e, 78.70.Dm, 77.80.Bh

# I. Introduction

The hexagonal manganites exhibit coupled ferromagnetism and ferroelectricity and hence are part of the class of multiferroic systems [1]. This hexagonal phase ($P6_3cm$) of $RMnO_3$ is found for small radius ions (RE=Ho, Er, Tm, Yb, Lu and also Y and Sc) [2,3,4], although the orthorhombic phase can be stabilized under wet chemistry conditions [5]. The region of hexagonal stability can be extended or reduced by deposition of $RMnO_3$ as films on strain inducing substrates [6]. From the applications perspective these materials have attracted much attention as media for data storage focused on use in nonvolatile random access memory (See Refs. [7] and references therein). These devices have the advantage over existing technology of low power consumption and decreased memory cell size because the electric charge induced by the remnant polarization controls the conductivity of the Si substrate onto which they are deposited. Low dielectric constant materials such as the hexagonal $RMnO_3$ systems ($YMnO_3$, $\varepsilon = 20$) do not suffer from the problem of high electrical noise due to the $SiO_2$ oxide at the interface. In addition, they do not posses toxic and volatile metals such as Pb and Bi. The hexagonal $RMnO_3$ materials satisfy these conditions and are being evaluated for use in devices. However, their properties in relation to their atomic and electronic structure are not well understood. A deep understanding is required to optimize these materials for use in data storage.

In this specific class of materials the transition to the ordered ferroelectric state ($T_{FE}$) occurs between ~800 and ~1200 K while the ordered magnetic states occur at significantly lower temperature ($T_N$~80 K). Anomalies in the dielectric constants, the linear expansion coefficients and phonon frequencies suggest a coupling between the magnetic and ferroelectric order at low temperature [8, 9]. In $HoMnO_3$, sharp features are observed at $T_{SR}$ and $T_R$, the in-plane spin rotation temperature and the R (R=rare earth) magnetic ordering temperatures, respectively. Recent, local structural measurements reveal changes in the bond lengths associated with the coupled magnetic sites which are responsible for the magnetic transitions [10]. Heat capacity and magnetic susceptibility measurements indicate that while the position of $T_N$ is stable with respect to external magnetic fields, $T_{SR}$ is strongly field dependent and moves to lower temperature with increasing magnetic fields [5]. Moreover, a complex multiphase structure is found at low temperatures. In magnetization studies on $HoMnO_3$ no slope changes are found at $T_N$. However, changes in slope were found at $T_{SR}$ and $T_{HO}$. The abrupt changes in the c-axis magnetization at $T_{SR}$ were attributed to changes in the magnetic order of the $Ho^{3+}$ ions since the Mn spin is confined to the a-b plane.

A fundamental understanding of the origin of ferroelectricity in these materials is needed in relation to the electronic and atomic structure. Lonaki *et al.* calculated the irreducible representations of the distortions from the high temperature symmetry group $P6_3/mmc$ to the low temperature symmetry group $P6_3cm$ [11]. This work revealed two phase transitions. Four different orthogonal modes were found in this analysis of which only one is ferroelectric. At high temperature the analysis suggested a polar to nonpolar transition occurs at which the $MnO_5$ polyhedra tilt away from being parallel to the z-axis. At lower temperatures a transition (from an intermediate antiferroelectric state) occurs at which the R planes buckles leading to the ferroelectric state.

Nenert *et. al.* conducted high temperature x-ray synchrotron powder diffraction measurement on $YMnO_3$ coupled with differential thermal analysis and dilatometery measurements on single crystals [12]. An argument was made for two phase transitions in this material one at 1100 K ($T_{c1}$) and one at 1350K ($T_{c2}$) with the latter being the transition to the high temperature paraelectric phase of P63/mmc symmetry. Their dilatometry measurements on unoriented single crystals suggested two transitions as cusps in the sample length vs. temperature. Differential thermal analysis on crushed single crystals also revealed two anomalies corresponding to transitions of unknown order. It was argued, based on the slow changes in the a and c lattice parameters between the two transition temperatures, that an intermediate phase exists between $T_{c1}$ and $T_{c2}$ and the changes at $T_{c1}$ and $T_{c2}$ are second order.

High temperature neutron diffraction measurements (sensitive to position of oxygen atoms) were conducted on $YMnO_3$ and $YbMnO_3$ between 1000 K and ~1400 K by Jeong *et al.* [13]. In these measurements it was claimed, based on the onset first order structural transition to a higher symmetry state, that the transition from the polar to nonpolar state occurs at 1200 K in $YMnO_3$. No such structural transition was found to occur up to 1350 K in $YbMnO_3$ and it was argued that the ferroelectric state persists up to this temperature. A more rapid reduction in tilting of the $MnO_5$ polyhedra is found in $YMnO_3$ which extrapolates to observed transition temperature using a power law dependence. The spread in the Y-O and Yb-O bond distances is found to decrease with increasing temperature. No evidence was found for a separate tilting followed by a FE transition. Interestingly, as in the case of $YbMnO_3$, our recent x-ray diffraction measurements on $ScMnO_3$ for temperatures up to 1300 K shown no first order structural transition [14].

Abrahams conducted a comprehensive study, by atomic coordinate analysis, of the $YMnO_3$ system [15]. Above $T_N$ the space group is argued to be $P6_3cm$ (ferroelectric), with a nonpolar paraelectric phase between $T_C$ and ~1360K. According to this work, the phase between $T_c$ and ~1360 K ($P6_3/mcm$) transforms to a second nonpolar paramagnetic phase ($P6_3/mmc$) with a reduction of cell volume by 3 which is stable up to 1600 K. Above this latter temperature it is predicted that the phase changes to (P6/mmm) a nonpolar paraelectric.

The most recent structural measurements over a broad temperature range by neutron diffraction structural measurements were carried out on $YMnO_3$ by Gibbs *et al.* [16]. In this work the authors observe the clear unit cell tippling transition near 1258 K with a change in space group from nonpolar $P6_3/mmc$ to polar $P6_3cm$ in detailed study of the high temperature structure. No change addition phase changes were observed on increasing to temperature up to 1403 K. In addition no intermediate phase is observed on cooling to room temperature. Evidence for an abrupt decrease in polarization based on a point charge model was found near ~ 920K. It is argued that the transition is isosymmetric and that the transition may related to an electronic transition involving a change in hybridization of the Y1-O3 bond. Fits to both $P6_3/mmc$ and $P6_3cm$ space groups were conducted with $P6_3cm$ giving the best fit.

Van Aken *et al*. argued that the mechanism behind ferroelectricity in $YMnO_3$ was driven by electrostatic and size effects rather than changes in chemical bonding [17]. It was also argued that ferroelectricity arises from buckling of Y planes stabilized by rotation of the $MnO_5$ polyhedra. The Born effective charges of all ions were found to be close to the formal values. It was suggested that the relative displacements of the ions are not driven by charge transfer or rehybridization of bonds and that there is no strong bonding between Y and O ions.

The electronic structure of the hexagonal phase of $RMnO_3$ differs significantly from that of the orthorhombic phase. For an $Mn^{3+}$ ion in octahedral symmetry the d states are split into a low lying $t_{2g}$ localized state and a higher energy $e_g$ state. The occupancy of the latter state for the $d^4$ configuration creates a Jahn-Teller instability in which the system energy can be lowered with the creation of a local distortion of the $MnO_6$ octahedra. For the hexagonal case, on the other hand, the $d^4$ Mn ion is in $D_{3h}$ symmetry which splits the Mn d levels into three components, one a1 and two $e_g$ states. Even in the high temperature state (high symmetry) the splitting between the a1 and eg states is large thus removing the degeneracy [18, 19, 20].

Cho *et al.* [20] investigated the electronic structure of the FE state of $YMnO_3$ on single crystals by examination of the Oxygen K-edge and Mn $L_{2,3}$ edge absorption spectra. The O K-edge spectra exhibit strong polarization dependence in the regions corresponding to the Mn 3d and Y 4d showing that strong Mn 3d –O 2p and Y 4d -O 2p hybridization exists. It is claimed that the change in polarization is due to rehybridization generated from the change in structure between the observed high temperature symmetric $P6_3/mmc$ structure and low temperature distorted $P6_3cm$ structure due to Y off centering.

First principles density functional calculations of the bandstructure of $YMnO_3$ were conducted for the high temperature paraelectric phase ($P6_3/mmc$) and low temperature ferroelectric phase ($P6_3cm$) for both the LSDA and LSDA+U approaches [21]. The electronic structure is not found to be sensitive to the crystal structure change in the two phases. The structural changes by themselves do not affect the band gap. Interestingly, it the on-site coulomb interaction parameter U plays an important role and shows the importance of electron correlation in the electronic structure of the $RMnO_3$ system. Independent of the structure, both antiferromagnetic order and a finite U parameter are required to produce the experimentally observed gap. The system is metallic for U=0. The onsite U separates the occupied and un-occupied 3d bands. More detailed noncolinear spin DFT calculations show that the nearest neighbor in-plane Mn-Mn spin frustration on the triangular lattice also plays an important in the formation of the band gap [22]. These calculations reveal that the lowest unoccupied band has $3d_z^2$-$O2p_z$ character. It was seen that the hybridization of the Mn3d and O2p is higher for the low temperature state.

In early electrical transport work on $YMnO_3$ single crystals, the pyoelectric current measured on warming after cooling in finite field is found to peak at 993 K indicating that this is this value corresponds to the ferroelectric transition temperature [23]. Measurement of the complex dielectric constant at high frequency (150 MHz) for single crystals of $RMnO_3$ R=Er, Ho, Y and Yb revealed a ferroelectric to paraelectric transitions at approximately 833 K, 873 K, 913 K and 993K, respectively [24]. Extensive electrical measurements were conducted recently [25]. Resistivity measurements on polycrystalline pellets of $HoMnO_3$ reveal breaks in the slope of the conductivity at 538 K and 873 K, possibly due to changes in the conduction mechanism and the ferroelectric-paraelectric transition. Changes in slope of the conductivity were found at 518 K and 933 K in $YMnO_3$ and at 973 K in $YbMnO_3$ for polycrystalline samples. In polycrystalline samples, the hysteresis loops corresponding to electrical polarization were found to vanish at approximately 973 K in $YMnO_3$ and 993 K in $YbMnO_3$. More recent resistivity measurements by Choi *et al.* on $YMnO_3$ reveal strong hysteresis for both in c-axis and ab-plane electrical resistivity measurements on single crystals in the region between 600 to 800 K showing that the onset of ferroelectricity coincides with changes in transport [26]. The transport measurements make it quite clear that $T_{FE}$ is near ~900K. All electrical measurements mentioned above are consistent with a continuous transition across $T_{FE}$. This raises the question about the true atomic origin of the ferroelectric transition.

In this work we explore the nature of the transition near ~900 K by carrying-out a detailed exploration of the structure of $RMnO_3$ on multiple length scales using x-ray spectroscopy and x-ray diffraction. X-ray absorption near edge (XANES) measurements of for both $YMnO_3$ and $HoMnO_3$ were conducted to examine the electronic structure changes (and hybridization changes). While a more complete set of measurements were conducted for $HoMnO_3$ including the extended fine structure vs temperature to probe the specific changes in bond lengths. An approach which makes few assumption such as imposing symmetry (local or long range) is required. Hence, for $HoMnO_3$, combined structural and ab initio molecular dynamics (AIMD) methods are utilized in this work.

## II. Experimental and Computational Methods

Polycrystalline samples of hexagonal of $YMnO_3$ and $HoMnO_3$ were prepared by solid state reaction. For measurements at or above 300 K, x-ray absorption samples were prepared by combining 500 mesh sieved powders with boron nitride and pressing the mixtures into pellets.

X-ray absorption local structure measurements covering the temperature range 6 K to 1038 K were performed to probe the electronic structure atomic structure. The low temperature $HoMnO_3$ data are taken from our previous work in Ref. [10]. The high temperature measurements (300K and above) were conducted at Brookhaven National Laboratory NSLS beamlines X11A and X19A. The limited energy range at the Mn K-edge constrained our modeling to the shells: <Mn-O1, Mn-O2, Mn-O3>, Mn-O4, Mn-Mn and Mn-Ho (long, short). For the Ho L3 edge, modeling was restricted to the <Ho-O>, Ho-Mn(short), Ho-Ho and Ho-Mn(long) distances as described in [10]. Data reduction and analysis follow that in Ref. [10], except that the fits at high temperature (above 300 K) involve co-refining the Mn K-edge and Ho L3 edge data (making the corresponding Mn-Ho and Ho-Mn distances equal). The data-range used for the Mn K-Edge was $2.55 < k < 11.3$ Å$^{-1}$ and that used for the Ho L3 edge was $2.94 < k < 14.0$ Å$^{-1}$. The $S_0^2 = 0.88$ value found at room temperature was used for all temperatures. The theoretical x-ray absorption near edge simulations follow Ref. [27]. To treat the atomic distribution functions on equal footing at all temperatures the spectra were modeled in R-space by optimizing the integral of the product of the radial distribution functions and theoretical spectra with respect to the measured spectra. Specifically the experimental spectrum is modeled by, $\chi(k) = \int \chi_{th}(k,r)\, 4\pi r^2 g(r) dr$ where $\chi_{th}$ is the theoretical spectrum and g(r) is the real space radial distribution function based on a sum of Gaussian functions ($\chi(k)$ is measured spectrum) [28] at each temperature. The positions and widths for Gaussian functions were optimized in the fits. Local structural measurements for YMnO3 we restricted to the XANES measurements in Fig. 2. High temperature x-ray diffraction measurements in quartz capillaries were carried out at NSLS X14A utilizing a Si strip detector (Fig. 3).

Calculation of the Born effective charges were base on spin density functional calculations in the projector augment wave approach [29] as in Ref. [10] utilizing $HoMnO_3$ structural parameters obtained from the diffraction experiments. For $YMnO_3$, neutron diffraction structural data were taken from Ref. [12]. To track the temperature changes in the structure with temperature, *ab initio* molecular dynamics (AIMD) simulations on $HoMnO_3$ with projected augmented wave methods and the LDA+U approximation with U/J = 8 eV/0.88 eV and (with the methods of Refs. [27]) were conducted in the (N V T) ensemble. A 2 x 2 x 1 unit (2a x 2b x c) supercell containing 120 atoms was used and P1 symmetry imposed to reduce assumptions based on symmetry constraints as done in Rietveld refinements in previous structural studies utilizing x-ray and neutron diffraction. A set of 9,000 time steps, each of 2 femtoseconds duration, were used for each chosen temperature. For each temperature run, the initial structure was taken as the room temperature experimental structure [30]. The last 8000 steps (16 picoseconds total time) were used in computing the average atomic positions in the cells shown in the structural figures below.

## III. Results and Discussion

In Fig. 1(a) we show the Mn K-edge x-ray absorption near edge spectra (XANES) in panel over the high temperature range for $HoMnO_3$. The spectrum is composed of the main line peak (first large peak) composed of "1s to "4p" transitions near 6555 eV. A low energy region near 6540 eV corresponds to 1s to "Mn np-hybridized with Mn 3d states" transition (np indicates a localized p state on Mn). This latter region enables one to view the Mn d-band indirectly and will be returned to below. The main peak broadens but exhibits no significant new features or splittings with increased temperature. We found the same behavior in $YMnO_3$ also over this temperature range and display it in Fig. 2.

For qualitative comparison (Fig. 1(b)) we calculated the XANES spectra for the high temperature phase (based on the $YMnO_3$ structure [12]) and the low temperature XANES spectrum. We also simulated the spectrum for a model in which the $MnO_5$ polyhedra are untilted (O1-Mn-O2 chain parallel to c, Fig. 1(c) ) but the Ho layers are buckled to simulate an intermediate phase. The simulations reveal that the main "4p" peak is broad and has two components for all of the calculations. Note that both the high temperature and the tentative intermediate phase simulation produced a low energy shoulder on the main line which is not seen in the measured data. This rules the possibility of an intermediate phase with a change of the $MnO_5$ polyhedral tilting only. We emphasize that these data indicate that no change in local symmetry is seen on crossing the ferroelectric transition temperature (~870 K in $HoMnO_3$ [25] or ~ 880 K in $YMnO_3$ [20]).

In Fig. 3 we show the $HoMnO_3$ lattice parameters as a function of temperature taken from our high temperature synchrotron x-ray diffraction data. Note the anomalous reduction in the c-axis length with increasing temperature which increase magnitude of the slope above ~900 K. This was also found in YMnO3 [12] and other $RMnO_3$ systems.

The local structure of $HoMnO_3$ as a function of temperature was determined by XAFS co-refinement of the Mn K-Edge and Ho L3-edge data. The structure functions are shown at both edges in Fig. 4. A continuous change with temperature is seen in the oxygen shell and higher shells with respect to both Mn and Ho sites. Note that the split Mn-Ho peaks is found over the entire temperature range measured. The detailed bond distances (Fig. 5(a)) as a function of temperature were extracted by fits to the x-ray absorption structure as done for the low temperature data following Ref. [10]. The distinct/split Mn-Ho distances (buckling of Ho layers) exist for all temperatures and show a trend toward merger at high temperature. The in-plane nearest neighbor <Mn-Mn> distance increases significantly at high temperature above $T_{FE}$. An important observation to note, however, is that the <Ho-O> distance shows an anomalous continuous decrease with at high temperature.

The temperature dependence of the atomic pair or bond correlations ("Debye-Waller" factors, $\sigma^2 = \langle(r - \langle r\rangle)^2\rangle$) were modeled by an static contribution ($\sigma_0^2$) plus a single parameter ($\theta_E$) Einstein model using the functional form $\sigma^2(T) = \sigma_0^2 + \frac{\hbar^2}{2\mu k_B \theta_E}\coth(\frac{\theta_E}{2T})$ [31] where μ is the reduced mass for the bond pair. This simple model represents the bond vibrations as harmonic oscillations of a single effective frequency proportional to $\theta_E$. It provides an approach to characterize the relative stiffness of the bonds. In Fig. 5(b) and 5(c) we show the Einstein model fits for <Ho-Mn> (long and short bonds), <Ho-O> and <Mn-Mn>. A large Einstein temperature is found for the <Ho-O> bond (~500 K) indicating strong bonding or strong Ho-O hybridization.

We found (Fig. 6) that in-plane and out-of-plane Mn-O distances increase systematically with temperature with no large anomaly but having large Einstein temperatures of ~900 K.

Under these assumptions in the model, examination of the Mn-O bond distances reveals no significant change on crossing the ferroelectric transition temperature. The bonds labeled <Mn-Oa> and <Mn-Ob> correspond to <Mn-O1, Mn-O2, Mn-O3> (mainly out-of-plane bonds, along c) and Mn-O4 (in-plane bonds, in ab plane). Note that the Einstein temperature (Fig. 6 (b)) is higher for the out-of-plane bond distribution, consistent with the large c-axis force constants which we found in Ref [10].

Based on x-ray diffraction data, density functional calculations (Table I) on $HoMnO_3$ and $YMnO_3$ were carried out and reveal strong hybridization of the oxygen atoms. However, no significant change in the Born effective charge tensor is found between the high temperature phases and the low temperature phase. The large asymmetry at O sites indicates strong hybridization of O with Ho and Mn.

In Fig. 7, we expand on the pre-edge region shown in Fig. 1(a). This pre-edge region is weak compared to the main line feature due to the fact that it is due to a transition from 1s to np state hybridized with the 3d band. The low energy side of this feature is seen to increase with temperature. This increase is indicative of increased density of unoccupied spin-up Mn $3d_{z2}$ – O 2p states which are the lowest bands found by DFT models. However, no abrupt change occurs with temperature. This is consistent with no large change in hybridization at the Mn sites with temperature (as also found in the Born effective charge calculations).

To obtain an understanding of the changes in atomic structure occurring in $T_{FE}$ without simplifying assumptions of the local or long range symmetry, we conducted *ab initio* molecular dynamics simulations on a 120 atom supercell of $HoMnO_3$ at 300, 500, 700, 900, 1100, 1250 and 1400 K in the lowest possible symmetry (P1). Figure 8 shows the displacements relative to the 300 K structure as arrows with respect to the center of atoms. The change at 900 K, where the polarization vanishes (Fig. 10 below), is dominated by z-axis displacement of the O4 and O3 ions and the Ho ions. The motion of O3/O4 ion results in a reduction of buckling of the MnO3/O4 planes. Examination of the structural parameters from AIMD simulations reveal that Ho-O distances in the 900 K data sets change by ~ 0.1 to 0.2 Å with respect to room temperature while the changes in the Mn-O distances are significantly smaller (~1/10 the magnitude of the changes in the Ho-O distances). The average Ho-O distance, however, is 2.28±0.01 Å at 300K and 2.31±0.01 Å at 900K. As temperature is increased, this buckling of the MnO3/O4 planes is further reduced as well as the buckling of the Ho planes with the additional rotation of the $MnO_5$ polyhedra so that the O1-Mn-O2 vectors move towards the z-axis. In figure 9 we show an expansion of the change in structure occurring at 900 K and (Fig. 9(a)). What should be noted is that the motions of the Ho, O3 and O4 ions result in movement of O3/O4 ions to central points between the Ho ions. It is this structural change that reduces the electric polarization. On moving to higher temperatures the buckling is further removed while preserving the increased symmetry obtained at lower temperatures. The final 1400 K structure is shown in Fig. 9(b) and 8(c). Note that it is at higher temperatures that rotation the polyhedral occurs and it is not related to a significant change in polarization.

Using the derived coordinates from the AIMD simulations the electrical polarization (computed in a simple point charge model using the formal charge of each ion) as a function of temperature is obtained in Fig. 10. The vanishing of the electrical polarization near ~900 K reveals that it is the Ho and O3/O4 z-axis motions enhancing the local symmetry at the Ho sites and not rotation of the $MnO_6$ polyhedra that are not responsible for the change in polarization. This is consistent with the local structural measurements. In addition the change is continuous starting from low temperatures as in the transport measurements. With respect to this unit cell averaged structure (like a diffraction measurement) the average Ho-O distance does not change significantly with temperature but the O coordination of Ho coordination become more symmetric above ~900K. These simulations suggest that in the real system, the unbuckling of the Mn-O3/O4 planes will result in an increased in-plane distances and a,b axis length while reducing the c axis length. This is exactly what is observed in Fig. 3.

The observed reduction of the Ho-O bond distance in the local structural measurements (x-ray absorption) can be understood by considering that at low temperature (below 900 K) the oxygen atoms between two Ho site on the z-axis are at two distinct distances leading to two signal which interfere resulting ina loss of both from the total XAFS signal. At high temperatures these bonds coincide in length and yield a reduction in the effective Ho-O bond length.

## IV. Summary

Measurements of the structure about the R and Mn sites were conducted in $RMnO_3$ for temperature significantly above the ferroelectric transition temperature ($T_{FE}$). The local and long range structural measurements were complemented by *ab initio* molecular dynamics simulations. With respect to the Mn sites, we find no large atomic (bond distances or thermal factors), hybridization changes or electronic structure changes on crossing $T_{FE}$ from local structural methods. With respect to the local structure about the Ho sites, a reduction of the average local Ho-O bond with increased temperature is found. Based on the density functional methods and diffraction data the Born effective charge tensor is found to be highly anisotropic at the O sites indicating very strong hybridization of the charge. The tensor does not change significantly above $T_{FE}$. *Ab initio* molecular dynamics calculations on $HoMnO_3$ reveal the detailed structural changes occurring with temperature. Above ~900 K there are large displacements of the Ho, O3 and O4 ions along the z-axis which reduce the buckling of the MnO3/O4 planes. It results in O3/O4 ions moving to central points between pairs Ho ions on the z-axis. These structural changes make the R sites more symmetric thus extinguishing the electric polarization. At significantly higher temperatures rotation of the $MnO_5$ polyhedra occurs without a significant change in electric polarization.

## V. Acknowledgments

This work is supported by DOE Grants DE-FG02-07ER46402 (NJIT) and DE-FG02-07ER46382 (Rutgers University). Data acquisition was performed at Brookhaven National Laboratory's National Synchrotron Light Source (NSLS) which is funded by the U. S. Department of Energy. This research utilized resources of the New York/Blue Supercomputer at the New York Center for Computational Sciences at Stony Brook University/Brookhaven National Laboratory which is supported by the U.S. Department of Energy under Contract No. DE-AC02-98CH10886 and by the State of New York.

**Table I. Born Effective Charge Tensor***

| Atom | $Z_{xx}$ | $Z_{xy}$ | $Z_{zz}$ |
|---|---|---|---|
| | **$HoMnO_3$ T=454K** | | |
| Mn | 3.25 | -0.01 | 3.88 |
| O1 | -1.95 | 0.00 | -3.25 |
| O2 | -1.95 | 0.00 | -3.28 |
| O3 | -3.06 | 0.12 | -1.61 |
| O4 | -3.14 | -0.04 | -1.41 |
| Ho1 | 3.67 | 0.00 | 4.15 |
| Ho2 | 3.80 | 0.00 | 4.13 |
| | **$HoMnO_3$ T=1133K** | | |
| Mn | 3.21 | -0.02 | 3.88 |
| O1 | -1.85 | 0.00 | -3.34 |
| O2 | -2.04 | 0.00 | -3.27 |
| O3 | -2.91 | 0.10 | -1.92 |
| O4 | -3.13 | -0.02 | -1.40 |
| Ho1 | 3.56 | 0.00 | 4.41 |
| Ho2 | 3.84 | 0.00 | 4.24 |
| | **$YMnO_3$ T=298K** | | |
| Mn | 3.19 | -0.01 | 3.83 |
| O1 | -1.99 | 0.00 | -3.14 |
| O2 | -1.91 | 0.00 | -3.26 |
| O3 | -3.02 | 0.12 | -1.54 |
| O4 | -3.05 | -0.04 | -1.42 |
| Y1 | 3.68 | 0.00 | 4.08 |
| Y2 | 3.75 | 0.00 | 4.00 |

* Units are in give in terms of the electron charge, e. Note that for this hexagonal system we have $Z_{xx} \approx Z_{yy}$, $Z_{xy} \approx Z_{yx}$ and $Z_{xz}=Z_{yz}=Z_{zx}=Z_{zy}=0.0$. The Born tensor obeys the acoustical sum rule.

# Figure Captions

**Fig. 1.** Measured temperature dependent XANES spectra of $HoMnO_3$ are shown in (a). The simulated XANES spectra for the high temperature (thin line), low temperature (thick line) and intermediate structure (no tilting of polyhedral but buckled Ho layers, dashed line) are given in (b). Panel (c) gives the low temperature crystal structure of hexagonal $HoMnO_3$ showing the tilted $MnO_5$ bipyramids .

**Fig. 2.** Measured temperature dependent XANES spectra of $YMnO_3$ revealing same trend as found in Fig. 2(a) for $HoMnO_3$.

**Fig. 3**. Temperature dependent lattice parameters of $HoMnO_3$ showing a change in slope occurring near ~900K as the possible onset of the ferroelectric state. Note that the c-axis is reduced with increased temperature.

**Fig. 4**. $HoMnO_3$ XAFS structure functions for temperatures from 298 K to 1038K for structure about the average Mn and Ho sites. Note that the buckling of the Ho layer is maintained for the entire temperature range studied.

**Fig. 5**. $HoMnO_3$ local structure derived (XAFS) temperature dependent bond distances are shown in (a) and the thermal factors with Einstein model fits are given in (b) and (c) for the <Mn-Mn>, <Ho-O>, <Ho-Mn>, and <Ho-Ho> distances.

**Fig. 6**. $HoMnO_3$ local structure derived (XAFS) temperature dependent bond distances about Mn are shown in (a) and the thermal factors with Einstein model fits are given in (b) for the <Mn-Oa> and <Mn-Ob> distances.

**Fig. 7**. The temperature dependence of the expanded pre-edge feature in $HoMnO_3$ ( for 1s to "Mn np-hybrizided with Mn 3d" region) shown in Fig. 1.

**Fig. 8**. Average structural changes in the 120 atoms $HoMnO_3$ super cell from *ab initio* molecular dynamics at temperatures of 500, 700, 900, 1100, 1250 and 1400K. The structure shown is for the 300K reference structure with arrows from atomic center indicating displacements from that structure at the given temperature. The red, purple and blue spheres correspond to O, Mn and Ho ions, respectively.

**Fig. 9**. Panel (a) gives change occurring in the 900 K structure manifested mainly as z-axis displacements of the O3, O4 and Ho ions. In (b) the 1400 K structure is shown with rotation of the polyhedral. The final structure at 1400 K is given in (c).

**Fig. 10**. Calculated $HoMnO_3$ polarization magnitude indicating that start to decrease at low temperatures and vanishes near ~900K, before the $MnO_6$ polyhedra rotate significantly and before the Ho planes loses their buckling. The polarization is along the z-axis.

**Fig. 1. Tyson *et al.***

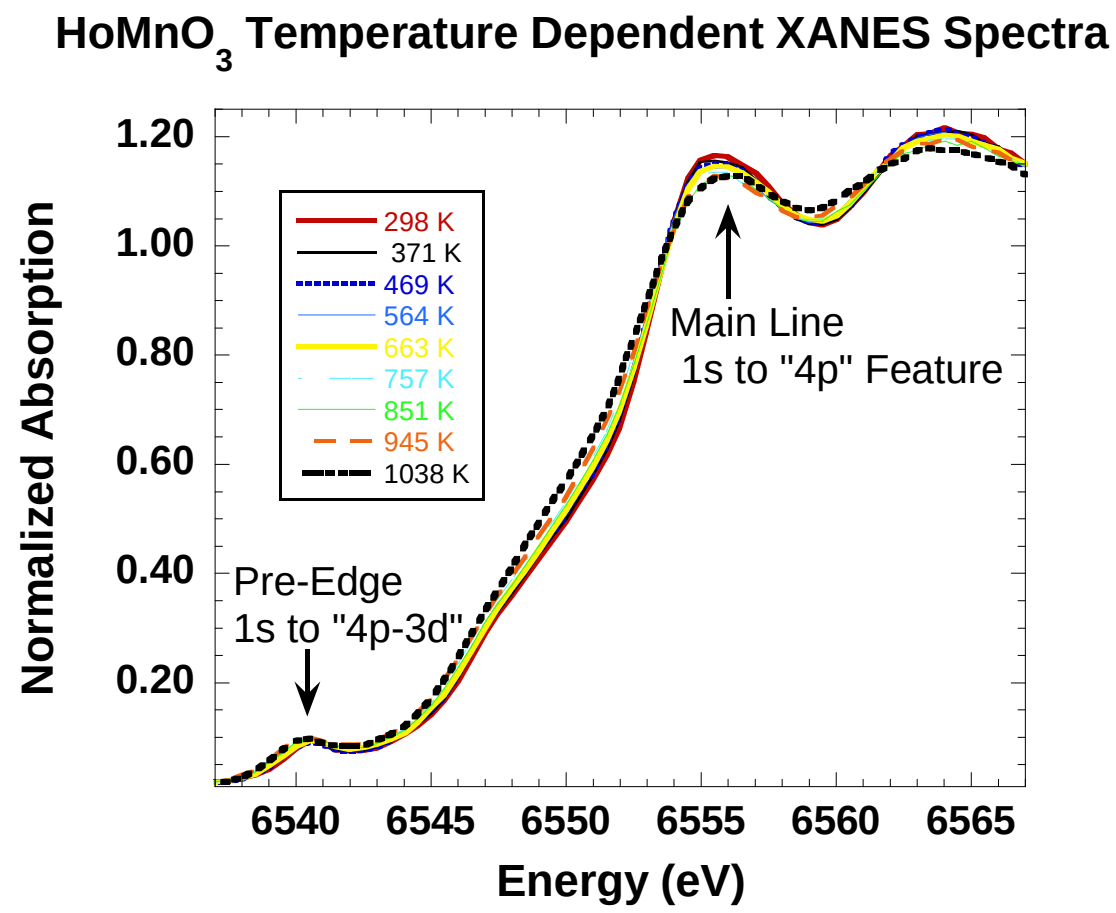

(a)

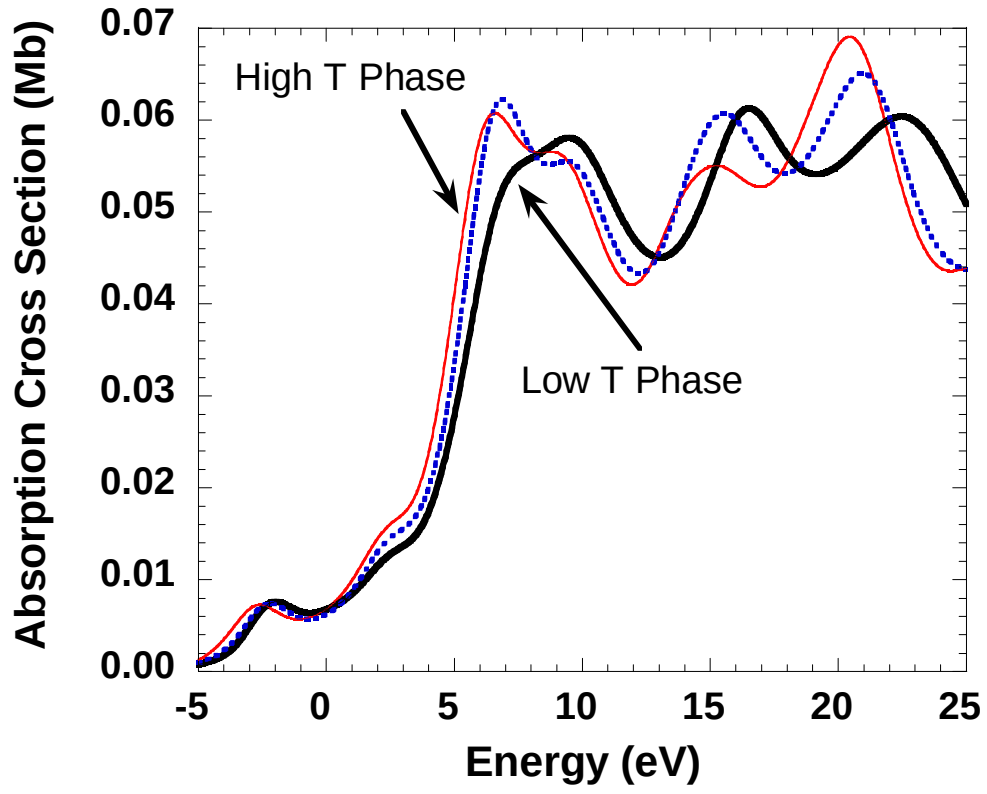

(b)

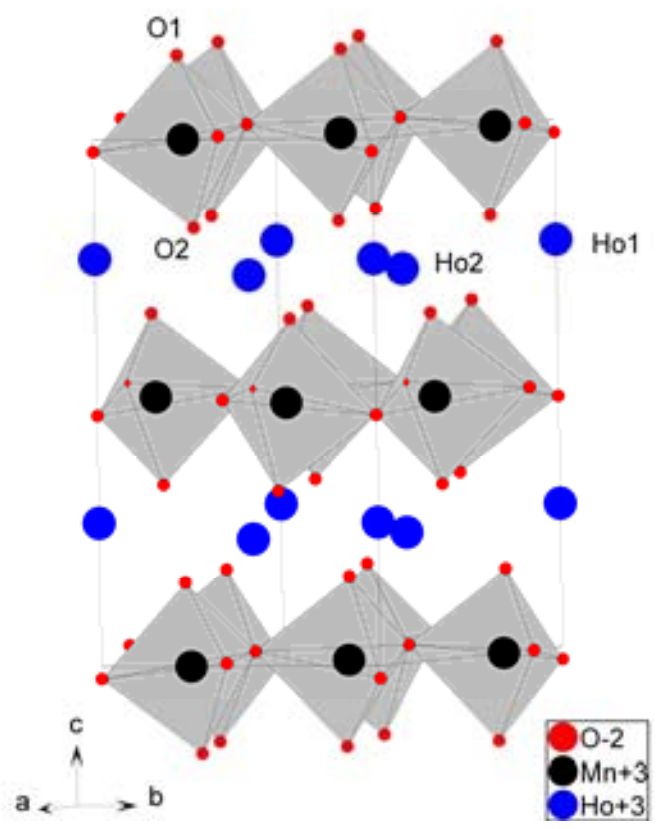

(c)

**Fig. 2. Tyson *et al.***

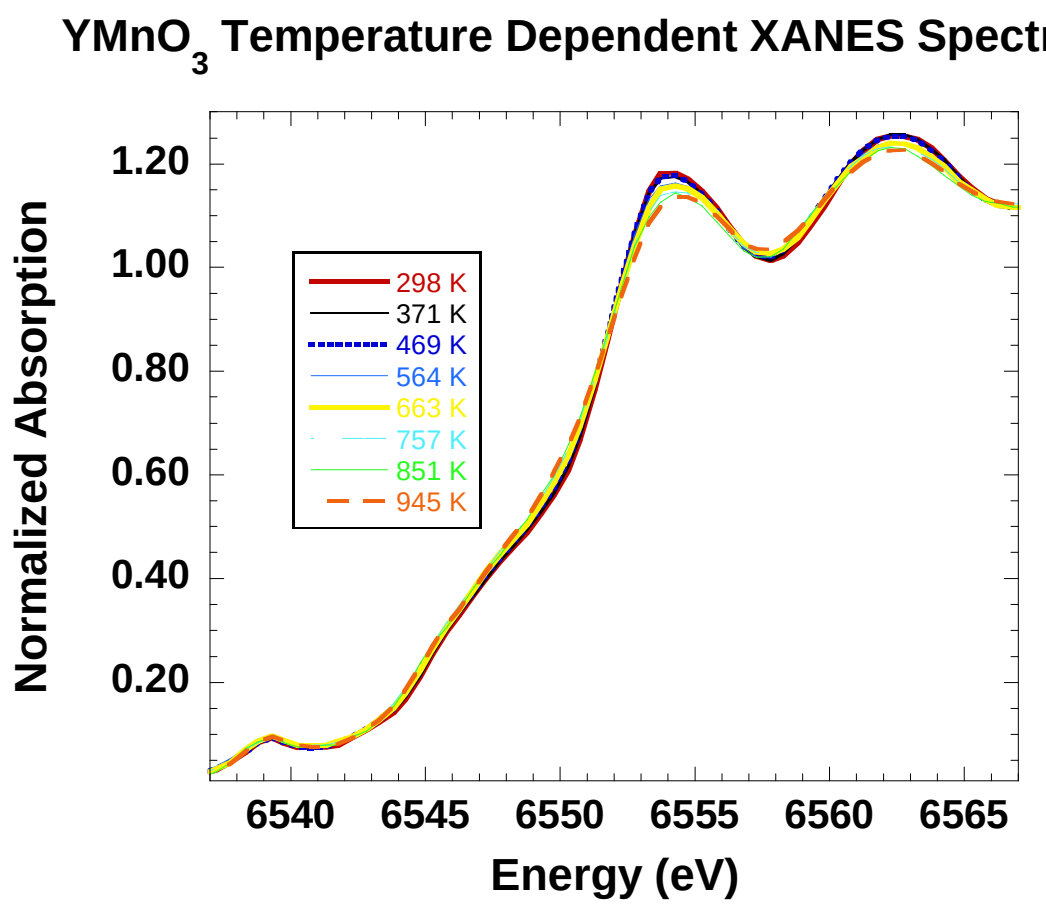

**Fig. 3. Tyson *et al.***

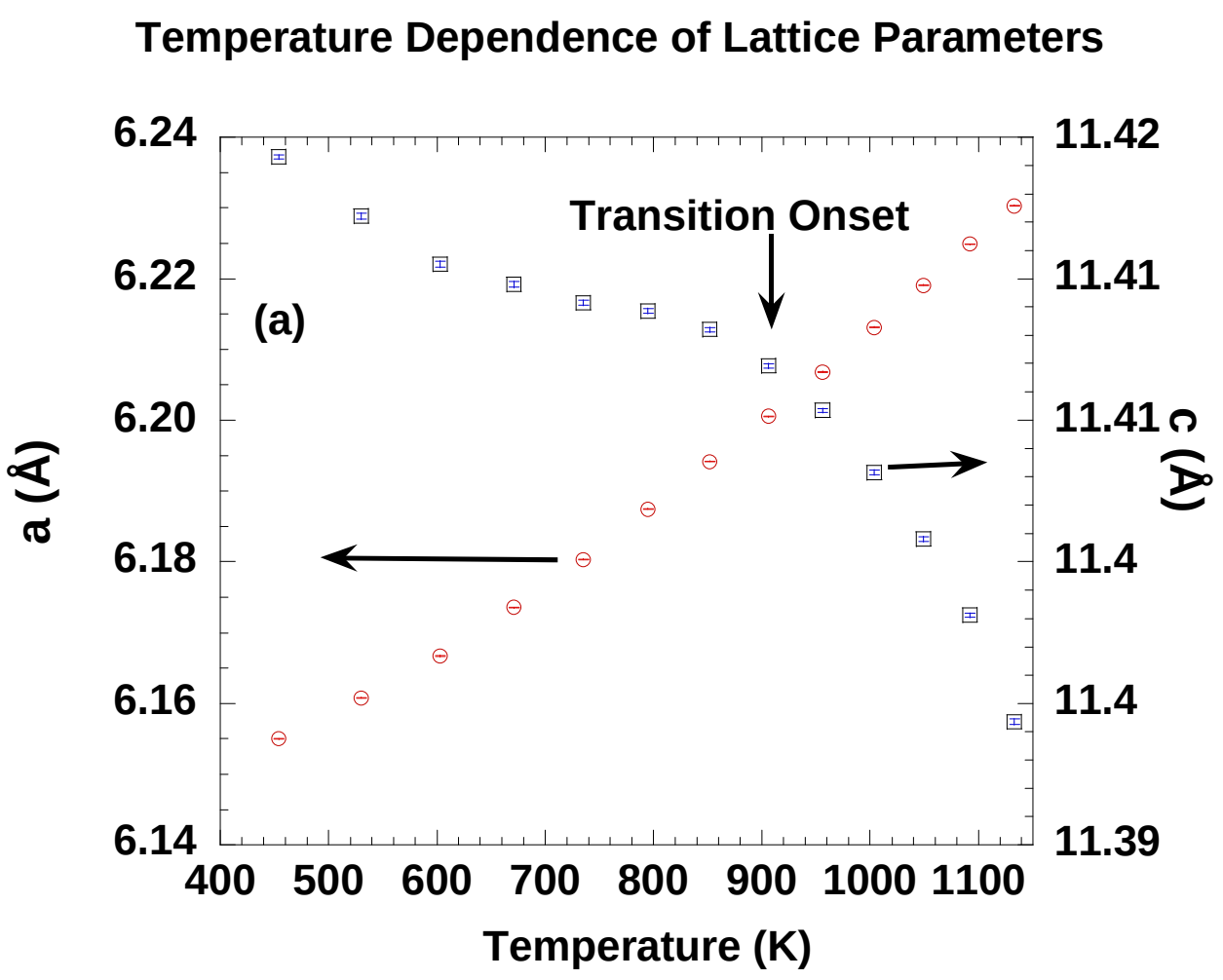

Fig. 4. Tyson *et al.*

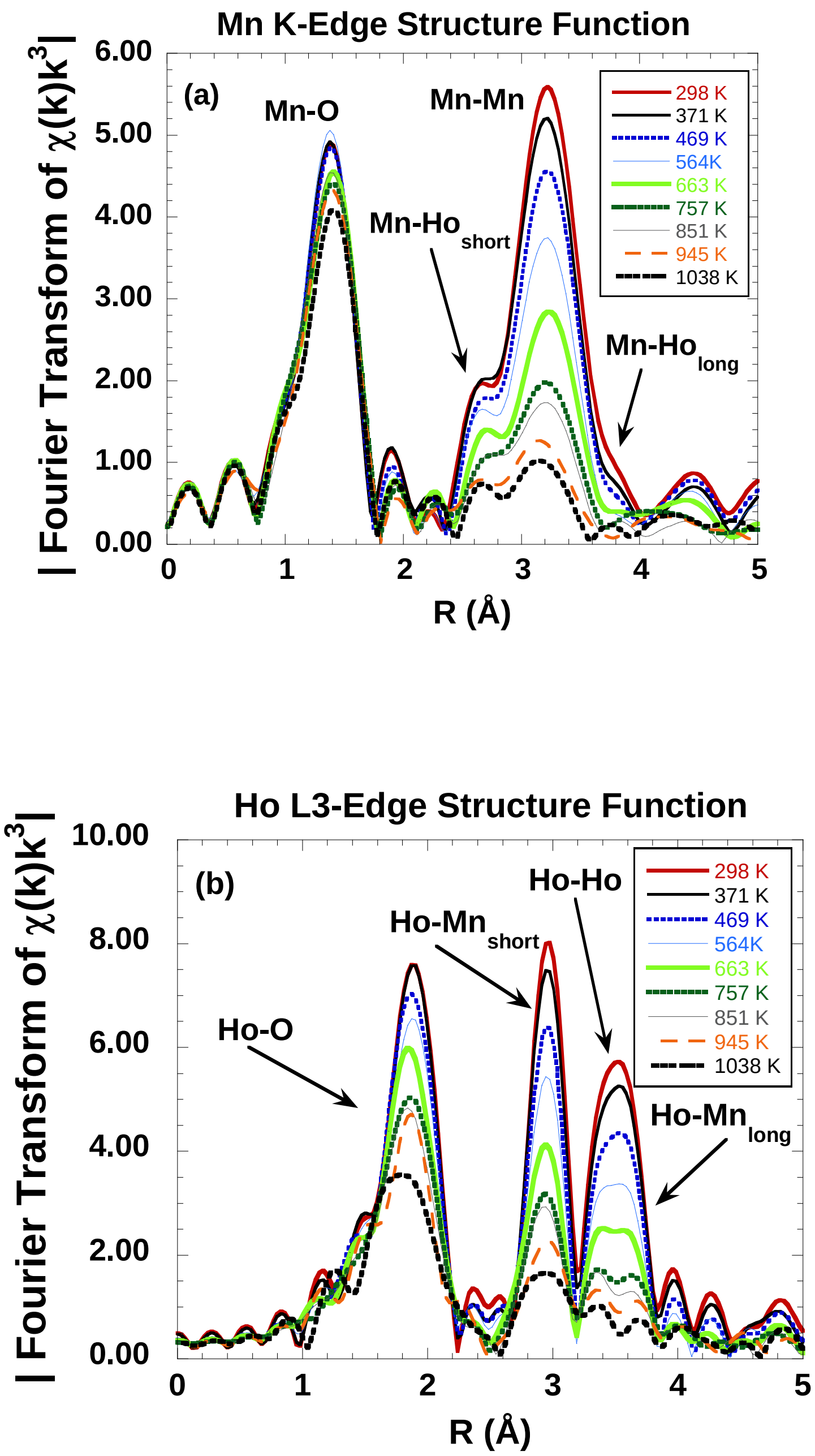

**Fig. 5. Tyson *et al.***

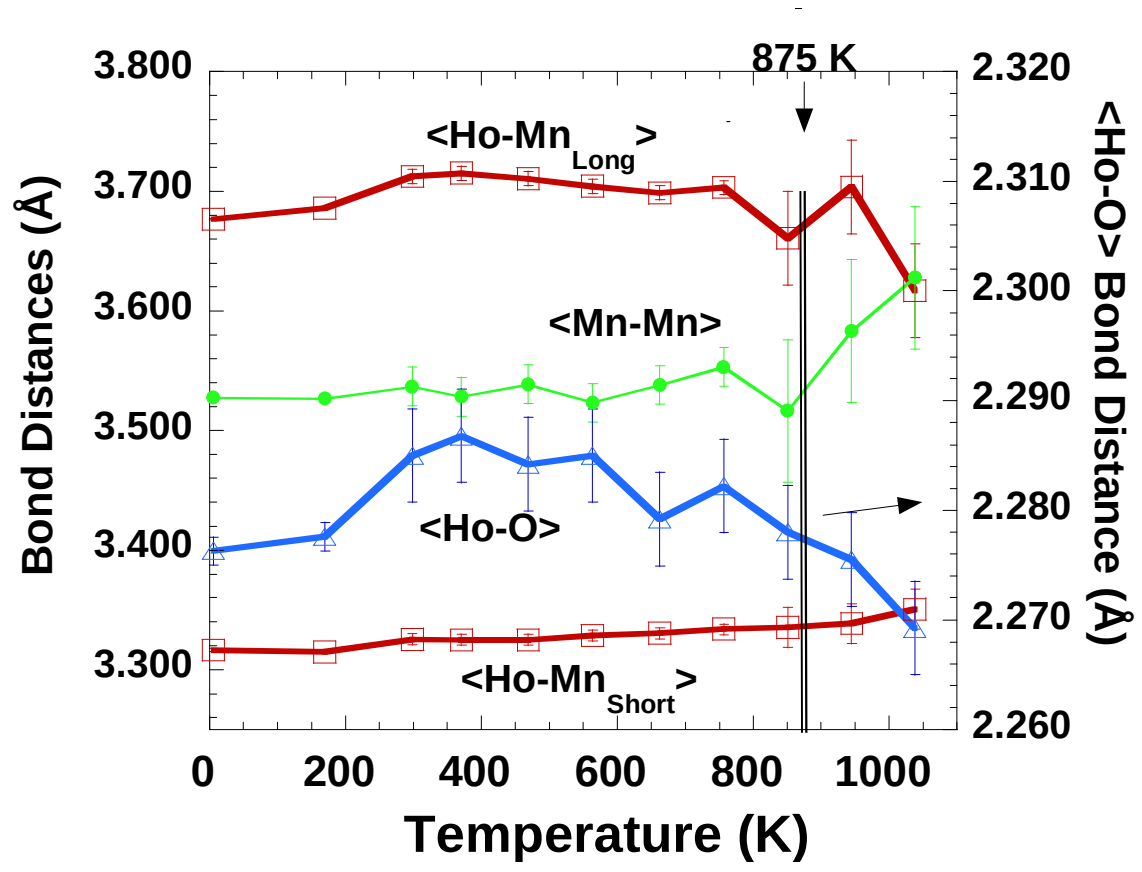

(a)

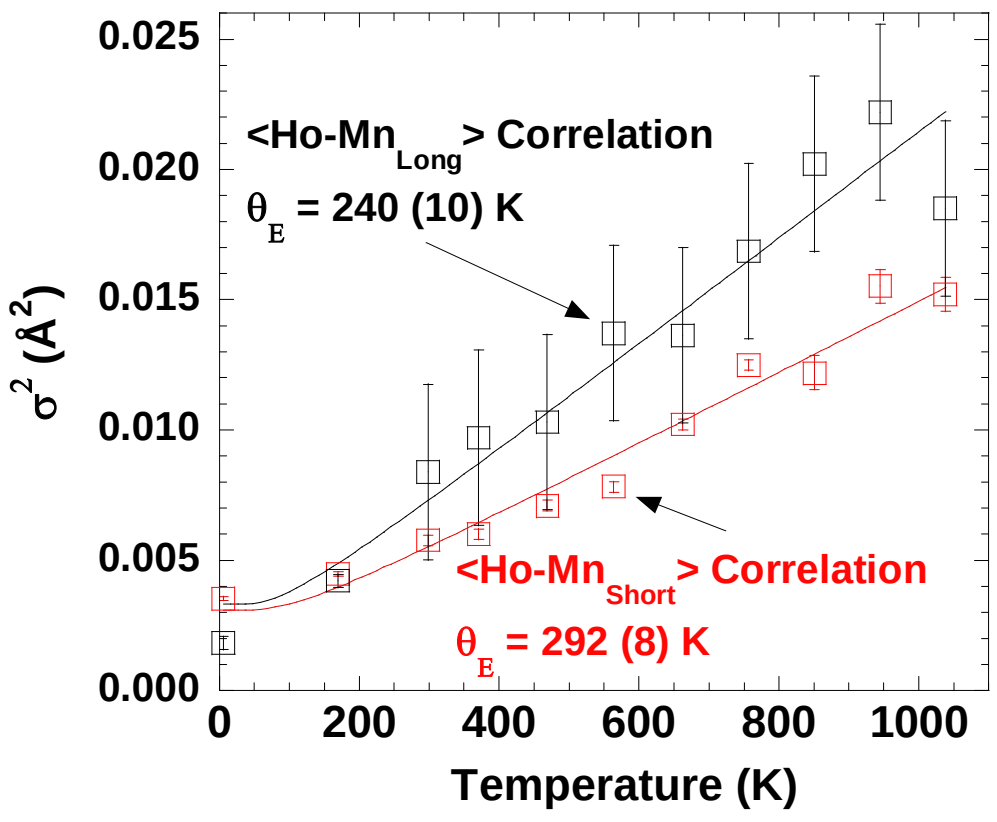

(b)

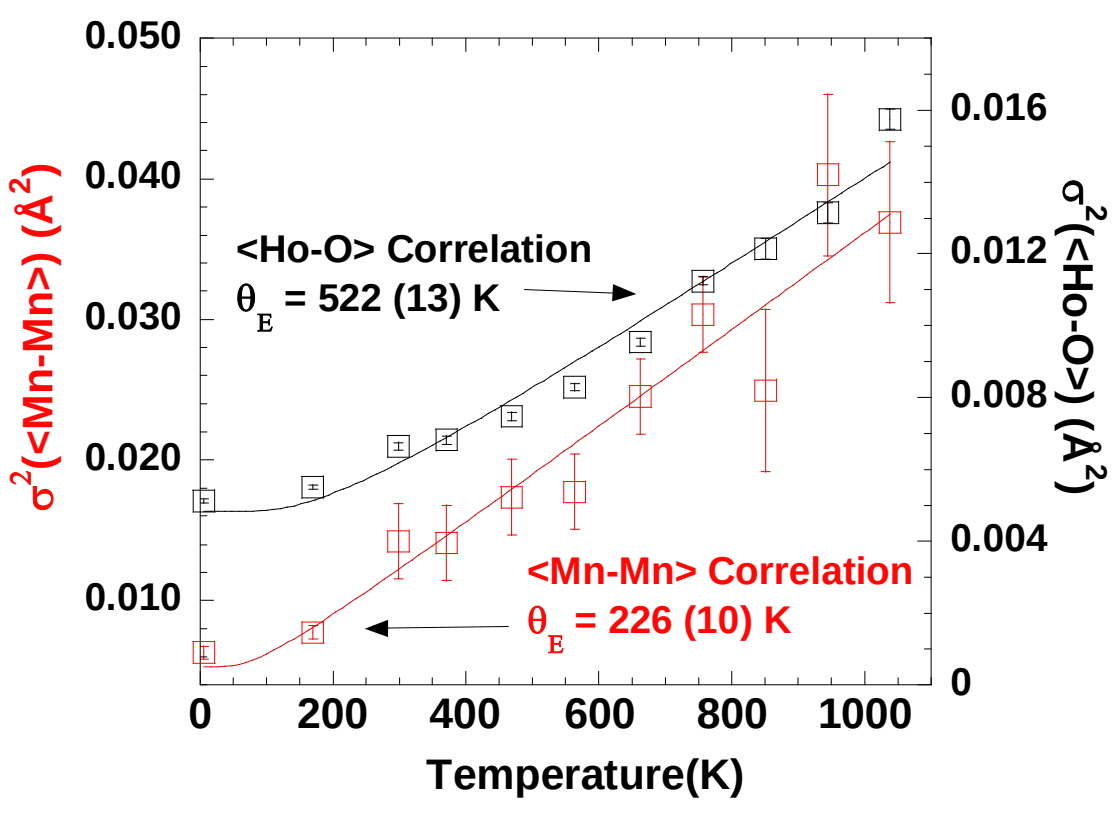

(c)

**Fig. 6. Tyson *et al.***

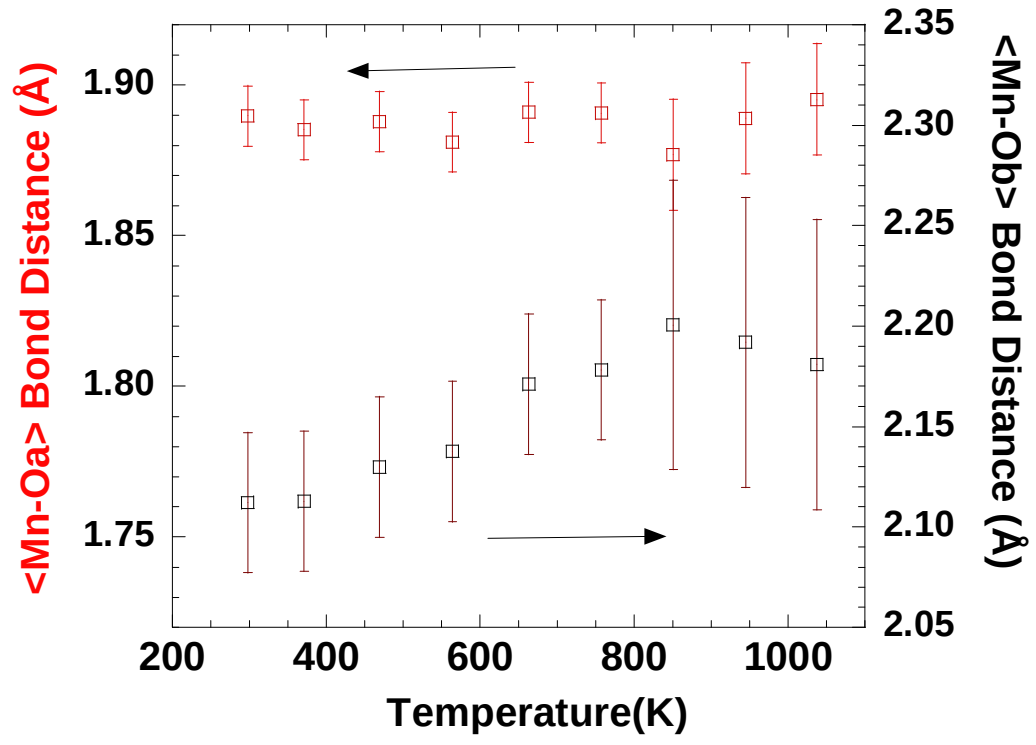

(a)

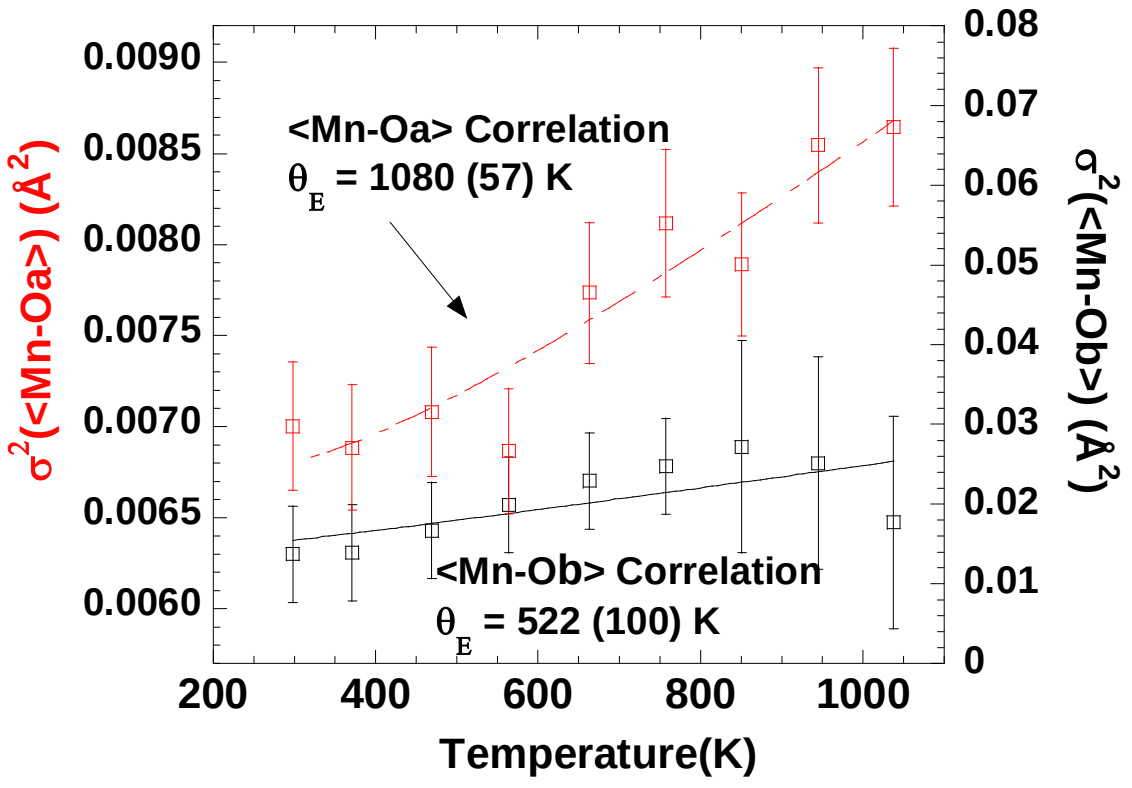

(b)

**Fig. 7. Tyson *et al.***

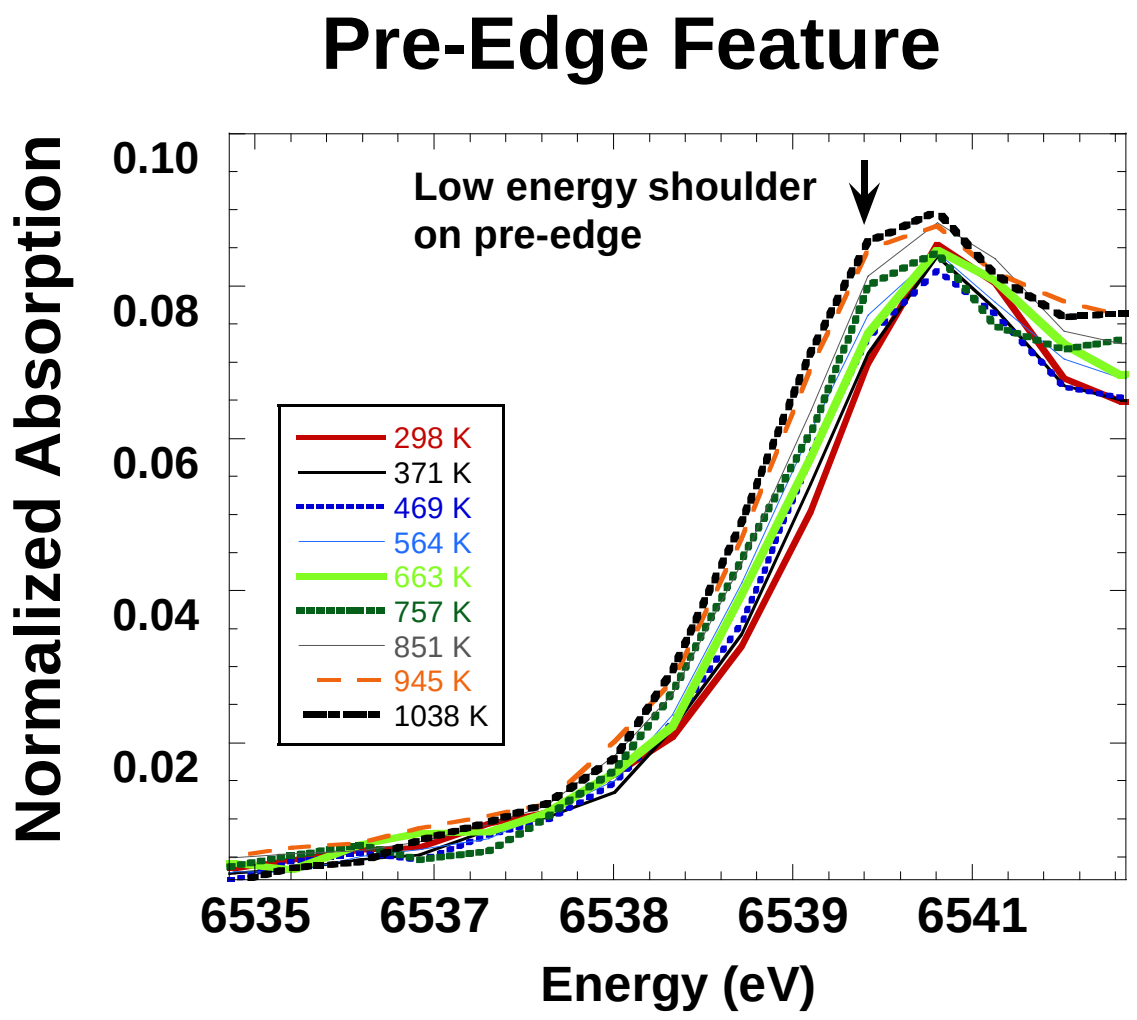

**Fig. 8. Tyson *et al.***

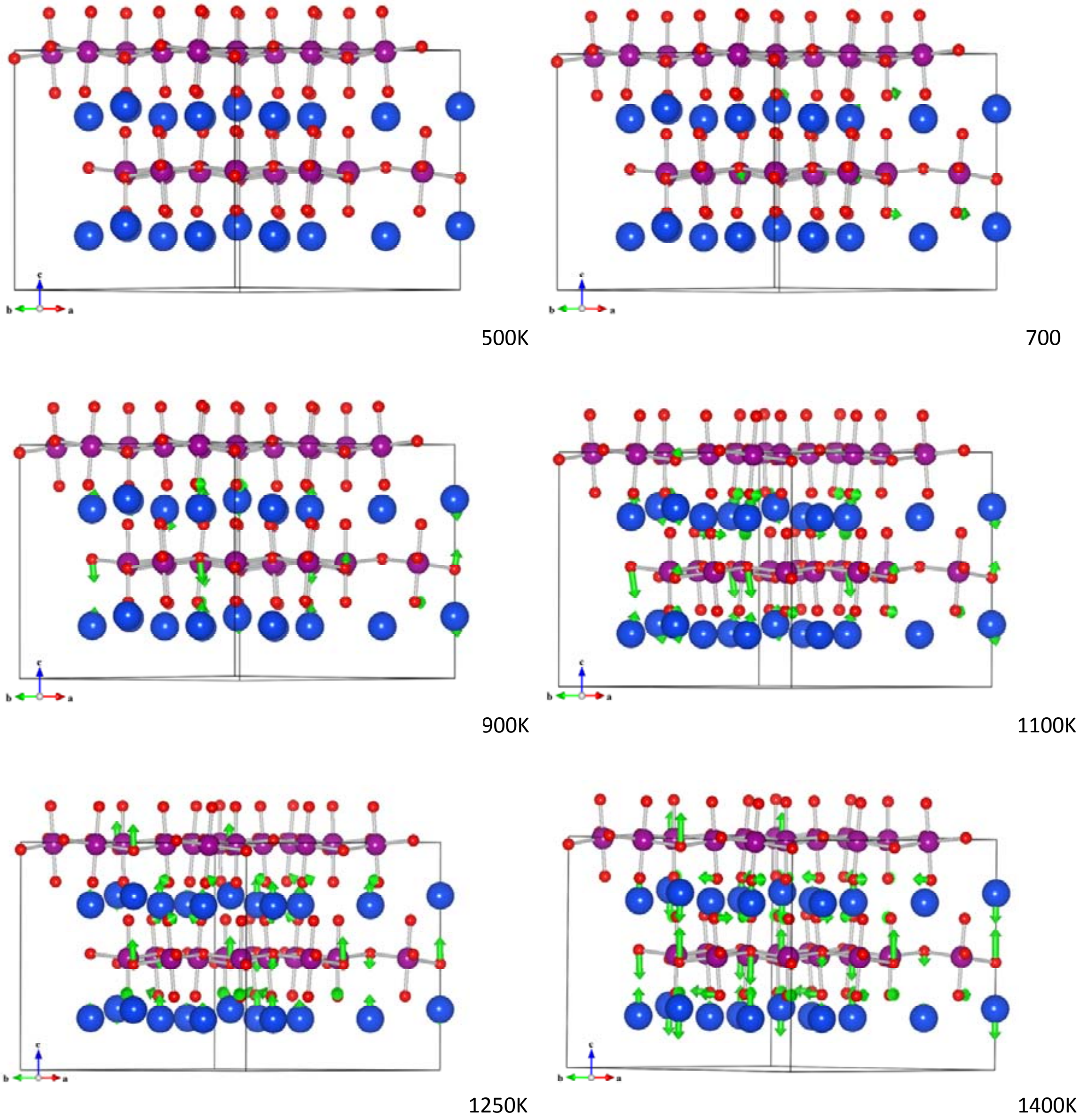

**Fig. 9. Tyson *et al.***

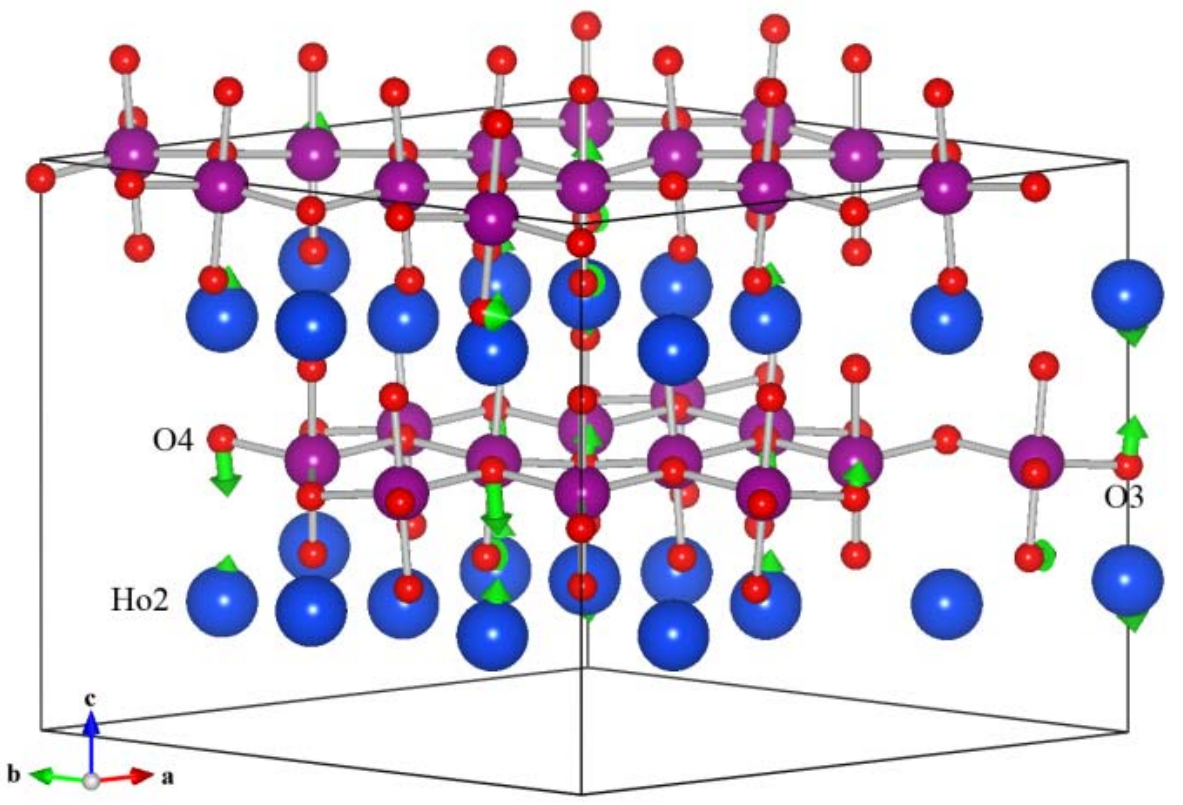

(a)

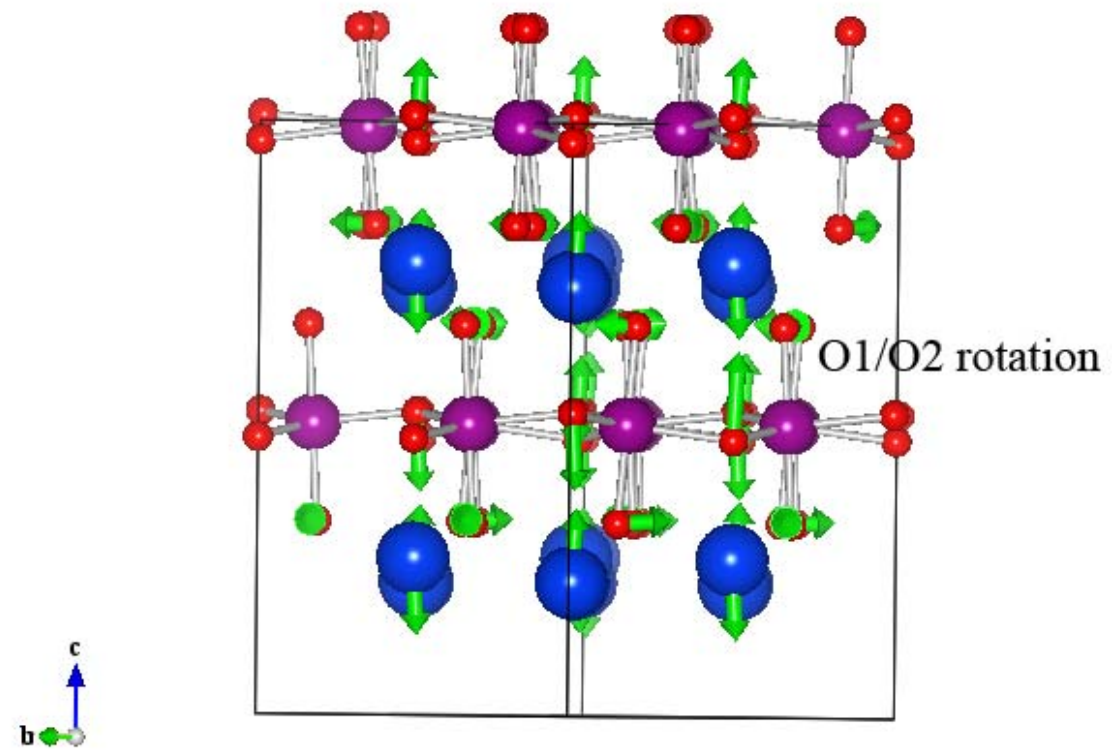

(b)

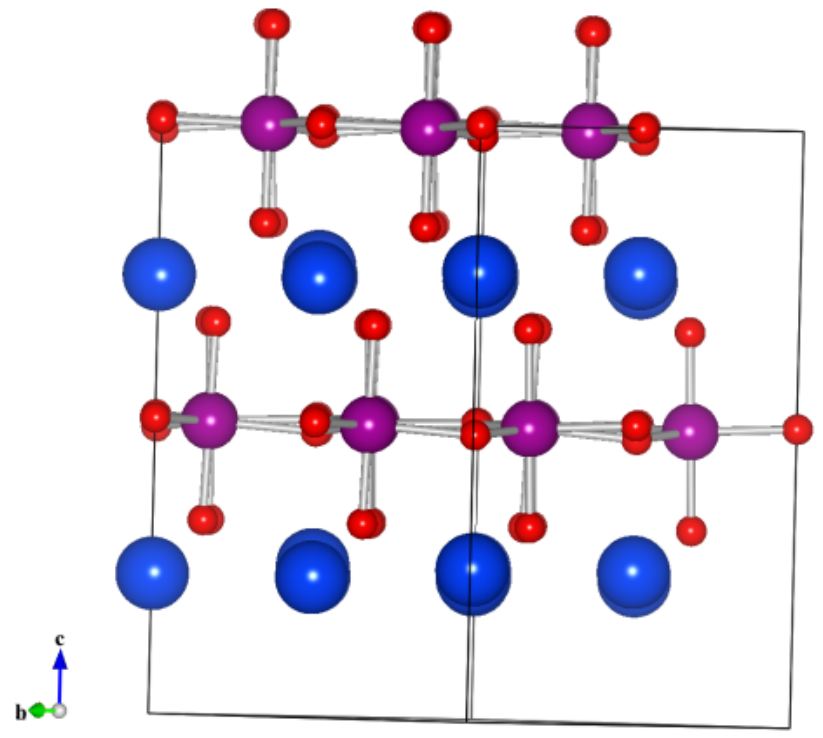

(c)

**Fig. 10.  Tyson *et al.***

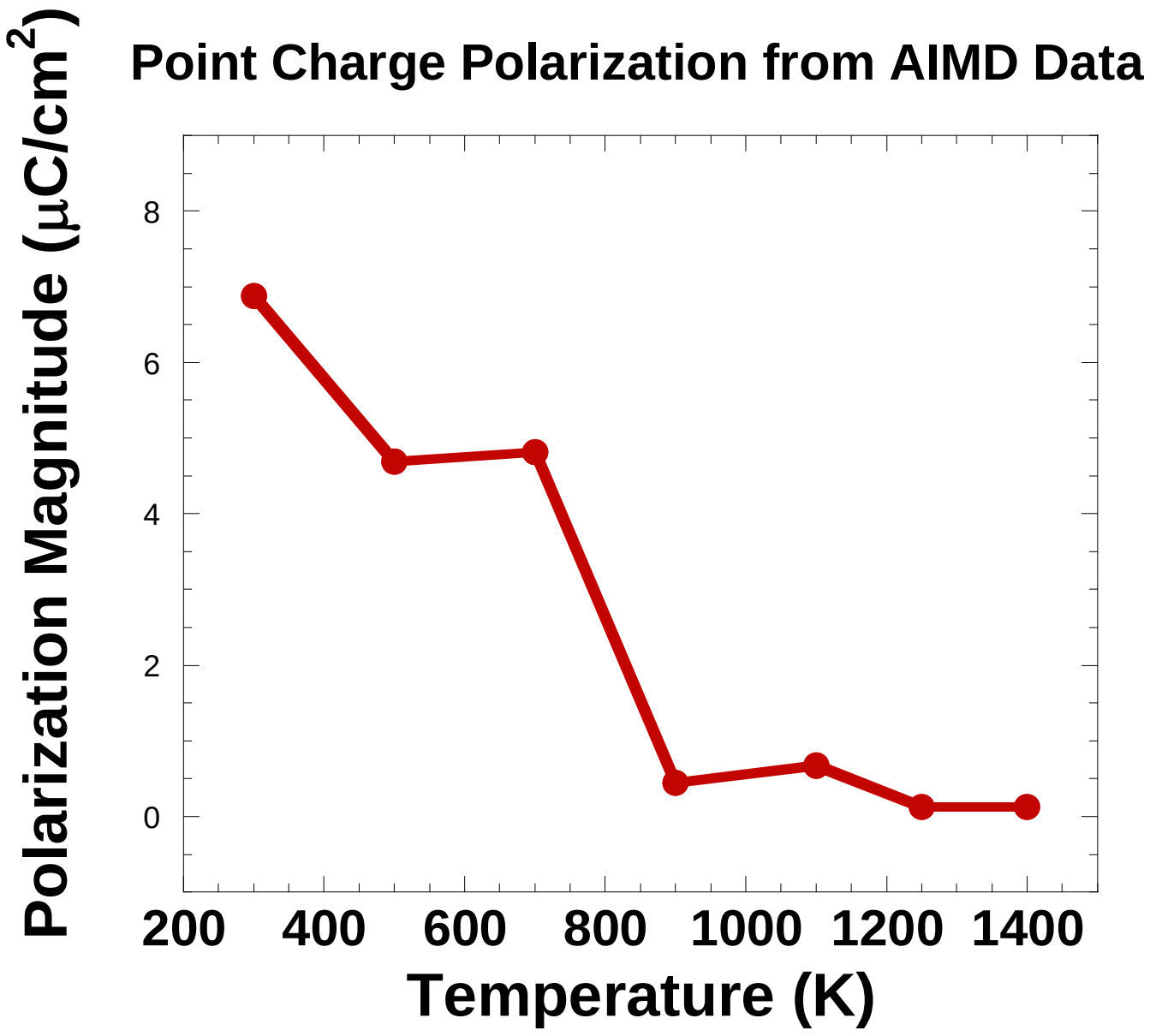